\documentclass{emulateapj}
\usepackage{natbib,epsfig}
\newcommand{\beq}{\begin{equation}}
\newcommand{\eeq}{\end{equation}}
\newcommand{\bea}{\begin{eqnarray}}
\newcommand{\eea}{\end{eqnarray}}

\newcommand{\rem}[1]{ }
\bibpunct[,]{(}{)}{;}{a}{}{,}

\shorttitle{Ion acceleration in shocks}
\shortauthors{Gargat\'e et al.}

\begin{document}

\title{Ion acceleration in non-relativistic astrophysical shocks}

\author{L. Gargat\'e \& A. Spitkovsky}
\affil{Department of Astrophysical Sciences, Princeton University, Princeton, NJ 08544-1001, USA}
\email{lgargate@astro.princeton.edu}

\begin{abstract}
We explore the physics of shock evolution and particle acceleration in non-relativistic collisionless shocks using multidimensional hybrid (kinetic ions/fluid electrons) simulations. We analyze a wide range of physical parameters relevant to the acceleration of cosmic rays (CRs) in astrophysical non-relativistic shock scenarios, such as in supernova remnant (SNR) shocks. We explore the evolution of the shock structure and particle acceleration efficiency as a function of Alfv\'enic Mach number and magnetic field inclination angle $\theta$. We show that there are fundamental differences between high and low Mach number shocks in terms of the electromagnetic turbulence generated in the pre-shock zone and downstream; dominant modes are resonant with the streaming CRs in the low Mach number regime, while both resonant and non-resonant modes are present for high Mach numbers.  Energetic power law tails for ions in the downstream plasma can account for up to $15\%$ of the incoming upstream flow energy, distributed over $\sim5\%$ of the particles in a power law with slope $-2\pm0.2$ in energy. Quasi-parallel shocks with $\theta\le45^\circ$ are good ion accelerators, while power-laws are greatly suppressed for quasi-perpendicular shocks, $\theta>45^\circ$. The efficiency of conversion of flow energy into the energy of accelerated particles peaks at $\theta=15^\circ$ to $30^\circ$ and $M_A=6$, and decreases for higher Mach numbers, down to $\sim2\%$ for $M_A=31$. Accelerated particles are produced by Diffusive Shock Acceleration (DSA) and by Shock Drift Acceleration (SDA) mechanisms, with the SDA contribution to the overall energy gain increasing with magnetic inclination. We also present a direct comparison between hybrid and fully kinetic particle-in-cell results at early times; the agreement between the two models justifies the use of hybrid simulations for longer-term shock evolution. In SNR shocks, particle acceleration will be significant for low Mach number quasi-parallel flows ($M_A < 30$, $\theta< 45$). This finding underscores the need for effective magnetic amplification mechanism in SNR shocks.

\end{abstract}


%
%
%
%
\section{Introduction}
\indent

Non-relativistic astrophysical shocks have long been considered as probable sources of acceleration of galactic cosmic rays (CRs) \citep{Chen:1975p1774,Bell:1978p884,Blandford:1978p891}. The CR spectrum extends over many decades in energy, following a power law $f(E)\propto E^{-\alpha}$ distribution. Nonthermal emission from supernova remnant shocks and from relativistic jets in Gamma-ray bursts and Active Galactic Nuclei is also modeled as synchrotron or inverse Compton scattering from a power law population of electrons at these sources. While the shape of the power law spectrum can be explained by modern shock acceleration theory, particle acceleration efficiency, i.e., the number and energy fraction contained in the accelerated particles, depends on nonlinear shock physics and has to be constrained through kinetic  simulations. Such simulations are the subject of this paper. 

The first-order Fermi acceleration mechanism at shocks \citep{Bell:1978p884,Blandford:1978p891,Blandford:1987p1773} is the strongest contender to explain the observed power law spectra. The mechanism works by having particles cross a shock front several times, with a positive energy gain on every crossing. Diffusive Shock Acceleration (DSA) theory describes how these particles scatter upstream and downstream from the shock due to electromagnetic (EM) turbulence, and predicts a universal power law slope $\alpha$, which to zeroth order depends only on the shock compression ratio ($n_d/n_u$, the ratio of the downstream to upstream density). Both the particle acceleration efficiency and the acceleration rate depend on the properties of turbulent fields and on the momentum of the accelerated particle; this is accounted for in the diffusion coefficient in the DSA model. Since the diffusion coefficient is not directly measurable in shocks of astrophysical origin, simplifying assumptions are usually made when estimating the maximum particle energies and acceleration times for a given object. One common assumption is to consider Bohm diffusion, by setting the diffusion coefficient to be proportional to the Larmor radius of the accelerated particle. This assumption then constrains the energy gain to be directly proportional to time, $E(t)\propto t$, and also limits the maximum energy of a particle by setting the maximum Larmor radius $r_{Li}$ to be of the same order of magnitude as the system size.  

Another relevant acceleration mechanism is Shock Drift Acceleration \citep[SDA,][]{Chen:1975p1774,Webb:1983p1775,Begelman:1990p1777} and Shock Surfing Acceleration \citep[SSA,][]{Lee:1996p3421,Hoshino:2002p3454,Shapiro:2003p3458}, where a particle is accelerated by the $\vec{E}_0=-\vec{V}\times\vec{B}$ upstream electric field in a magnetized shock, with $\vec{B}$ the upstream magnetic field, and $\vec{V}$ the upstream plasma flow velocity measured in the downstream frame. The main difference between SDA and SSA is the importance of the electrostatic cross-shock potential for the confinement of particles in the acceleration region; nevertheless, particles are accelerated by the same upstream electric field $E_0$, and for the purposes of this work we do not distinguish between the two. After being reflected at the shock, a particle will drift along the shock front in the direction of this upstream electric field, gaining energy. For magnetized shocks, this mechanism works along with DSA to produce energetic particles.

Astrophysical collionless shocks can be characterized by their Alfv\'enic Mach number, defined as $M_A=V_{sh}/V_{A}$, where $V_{sh}$ is the shock front velocity (in the upstream frame) and $V_A=B_0/\sqrt{4 \pi n_0 m}$ is the Alfv\'{e}n velocity (with $B_0$, $n_0$ and $m$ being the upstream magnetic field, ion number density and  ion mass, respectively). The magnetic field inclination angle $\theta$ between $\vec{B}$ and the shock normal also plays an important role. Generation of electromagnetic turbulence in the pre-shock zone is crucial for the DSA mechanism, since particles need to be scattered back to the shock front multiple times for efficient acceleration. Since cold particles will be constrained to move along the magnetic field lines, different shock configurations will develop for angles $\theta\le 45^\circ$ (quasi-parallel shocks), and for angles $\theta>45^\circ$ (quasi-perpendicular shocks). The main difference for the two cases is that for $\theta>45^\circ$ there is a sharp decrease in the number of particles escaping from the shock. For quasi-parallel shocks an extended pre-shock region will develop, with waves and turbulence being generated by reflected ions that stream in front of the shock; quasi-perpendicular shocks will exhibit very little self-generated turbulence upstream, and instead a shock-foot region with the size comparable to the ion Larmor radius $r_{Li}$ will form. The DSA mechanism will then play a crucial role in particle acceleration in quasi-parallel shocks \citep[for a comprehensive review of microphysics of collisionless shocks see][]{Treumann:2009p2090}.

Several approaches are commonly used to model the intrinsically nonlinear processes that dominate shock acceleration. Semi-analytic kinetic models \citep{Kirk:1989p1778,Ballard:1991p1780,Achterberg:2001p1795} provide a theoretical background for the description of the particle acceleration problem. Similarly, Monte Carlo test particle simulations with prescribed turbulence models have been performed by several authors \citep[e.g.,][]{Bednarz:1998p1798,Ellison:2004p1800,Niemiec:2004p1799}. These can be extended over very large periods of time and thus have predictive behavior over the long term evolution of particle acceleration. They lack the self-consistent properties of first principle simulations, such as fully kinetic particle-in-cell (PIC) models. In PIC models, the full set of Maxwell's equations is solved on a finite grid, where particles are also pushed and serve as source terms for the creation of EM fields \citep{birdsall}. Hybrid simulations follow a similar method and consider kinetic ions and fluid electrons \citep{lipatov}; since electron time-scales do not need to be resolved, evolution of the physical system for longer times is possible. 

Previous work on multidimensional PIC shock simulations has been conducted for relativistic pair plasmas \citep{Spitkovsky:2005p1892,Spitkovsky:2008p1909,Sironi:2009p1330,Nishikawa:2009p3515}, and for relativistic ion-electron plasmas \citep{Spitkovsky:2008p1910,Martins:2009p1933,Sironi:2011p3059}. In the non-relativistic regime, both PIC \citep{Amano:2007p2166,Kato:2008p2139,Amano:2009p2160,Amano:2010p2162,Riquelme:2011p3505} and hybrid simulations exist \citep{Bennett:1995p1175,ELLISON:1993p40,GIACALONE:1992p45,GIACALONE:1994p41,GIACALONE:1997p1192,SCHOLER:2002p2491,Giacalone:2004p44,Sugiyama:2011p3494}, among others. Electron acceleration is found to be more efficient in (magnetized) quasi-perpendicular shocks, in the non-relativistic case, while ion acceleration is mainly observed in quasi-parallel shocks.

In this work, we concentrate on the systematic analysis of non-relativistic magnetized ion-electron shocks for diverse Mach numbers and magnetic field inclination angles. We extend previous work by analyzing a wide range of parameters, relevant for both solar wind shocks and for SNR shocks, and follow the shock evolution and ion acceleration for very long times and extended regions of space in two dimensions. Such extensive surveys are not commonly found in the literature (\citet{GIACALONE:1997p1192} presents a similar survey, done in 1D, and for a more limited range of Mach numbers). Qualitative differences are found between high and low Mach number shocks, and for quasi-parallel and quasi-perpendicular shocks. The analysis of individual ion trajectories provides a direct measurement of how particles are accelerated, and allows us to discriminate between the DSA and SDA models, and their relative contributions to the high-energy end of the particle distribution functions. We focus on particle acceleration and how the main shock parameters affect the acceleration efficiency. We also analyze the particle spectra behavior in time, and compare between spectra for the different Mach numbers and B field inclination angles, providing a complete survey of ion acceleration in non-relativistic magnetized shocks.

This work is organized as follows. In Section \ref{section2} we present the simulation setup and the physical parameters to be explored. In Section \ref{section3} we show the global shock properties, including the main features of quasi-parallel and quasi-perpendicular shocks, together with typical particle spectra found in each case. In Section \ref{section4} we present the results of the complete shock parameter survey, analyze the shock properties with particular emphasis on wave and turbulence generation, and its influence on the measured particle spectra. In Section \ref{section5} we consider the particle acceleration mechanisms and discuss how their relative contributions vary with the shock parameters. The final discussion, consequences and constraints for astrophysical shock scenarios derived from the present work are presented in Section \ref{section6}. Finally, we add an Appendix presenting a direct comparison between fully kinetic PIC simulations and their hybrid counterparts, and we include a detailed analysis of the numerical parameters used.

%
%
%
%
\section{Shock setup and simulation parameters}
\label{section2}

Two particle simulation codes are used to produce the results shown throughout this work: most results are obtained with the kinetic ion/fluid electron code \textit{dHybrid} \citep{Gargate:2007p117}, and then checked with the relativistic fully kinetic PIC code \textit{TRISTAN-MP} \citep{buneman,Spitkovsky:2005p1892}. While both codes are three-dimensional, in this work we only use them in two-dimensional configuration. The hybrid and PIC simulation models are very similar at the core. Both codes use a discrete grid where the electric and magnetic fields are calculated, which are used to advance particle velocities and positions in time; charges and currents from the particle populations are then used as source terms in Maxwell's equations to re-calculate fields. The hybrid approach differs from PIC in that the displacement current is neglected in Amp\`ere's law, i.e.,$\nabla\times\vec{B}=4 \pi \vec{J}/c$, eliminating the propagation of light waves, and the electrons are modeled as a massless charge-neutralizing fluid. The electron equation of state is such that, after being adiabatically heated at the shock, the electron distribution downstream will account for $\sim50\%$ of the upstream ion flow energy; such a choice is physically motivated by results from PIC simulations. Details on the general properties of hybrid codes can be found in \citet{lipatov} and of PIC codes in \citet{birdsall}.
 
In order to simulate a shock we consider an in-plane $B$ field, propagate plasma through our simulation box in the $-x$ direction, and reflect the flow from the left simulation wall. The two counter-propagating flows that result are electromagnetically unstable. The two flows thermalize and a shock propagating in the $+x$ direction is produced; the simulation frame is the downstream plasma frame for this setup. For electron-proton magnetized shocks, which are the topic of this work, the initial plasma flow velocity $\vec{V}$, the proton (or electron) number density $n$ and the background magnetic field intensity and direction completely determine the physical properties of the shock. For typical astrophysical shocks the inflowing upstream plasma is cold, such that the magnetosonic Mach number is similar to the Alfv\`enic Mach number, and a strong (in the MHD sense) shock will form if $M_A=V/V_A\gg1$.
%
\begin{deluxetable}{ccccc}
\tablecaption{Key simulation parameters\label{table1}}
\tablehead{\colhead{Run}&\colhead{$M_A$}&\colhead{$\theta$}&\colhead{Box size ($\mathrm{c/\omega_{pi}}$)}}
\startdata
$\mathcal{A}_{1...6}$&3.1&0,15,30,45,60,75&$2000\times100$\\
$\mathcal{B}_{1...5}$&6.0&0,15,30,45,60&$2000\times100$\\
$\mathcal{C}_{1...6}$&10&0,15,30,45,60,75&$5000\times100$\\
$\mathcal{D}_{1...5}$&31&0,15,30,45,60&$8000\times100$\\
$\mathcal{E}_{1...4}$&100&0,15,30,45&$8000\times100$\\
$\mathcal{F}$ (PIC)&4.7&0&$296\times18$\\
$\mathcal{G}_1$&4.7&0&$296\times18$\\
$\mathcal{G}_2$&4.7&0&$2000\times18$\\
$\mathcal{G}_3$&4.7&0&$2000\times100$\\
$\mathcal{G}_4$&4.7&0&$2000\times4$\\
\enddata
\end{deluxetable}

The results are presented in normalized units, with the magnetic field normalized to the upstream background field $B_0$, the density normalized to the upstream plasma density $n_0$, the velocities normalized to the Alfv\'en velocity, the spatial dimensions normalized to the ion inertial length $c/\omega_{pi}$ (with $\omega_{pi}=(4 \pi n_0 e^2/m)^{1/2}$ the ion plasma frequency, and $c$ the speed of light), and time normalized to the inverse ion cyclotron frequency $\omega_{ci}^{-1}=(e B_0/m c)^{-1}$. For hybrid runs, the grid cell size is set to $0.2\,c/\omega_{pi}$, the time step is set to $\Delta t=0.01\,\mathrm{\omega_{ci}^{-1}}$, and 4 particles per cell are used; total simulation times of up to $1200\,\mathrm{\omega_{ci}^{-1}}$ were considered. For PIC runs, we use a grid cell resolution of $0.067\,\mathrm{c/\omega_p}$ (electron skin-depths), and 4 particles per cell; a reduced $m_i/m_e=30$ mass ratio is employed, in order to improve computation time efficiency.

The full set of our runs is summarized in Table \ref{table1}. Simulation box sizes vary with Mach number, from $2000\times100\,\mathrm{(c/\omega_{pi})^2}$ up to $8000\times100\,\mathrm{(c/\omega_{pi})^2}$, for most of the hybrid runs. In the runs from sets $\mathcal{A}$ through $\mathcal{E}$, a thorough physical parameter survey was conducted, including shock Mach numbers from $3.1$ through $100$, and $\vec{B}$ field inclination angles from $0^\circ$ through $75^\circ$, as specified in the table. In the following sections we thoroughly analyze the shock properties, particle spectra, and acceleration mechanisms observed, for runs in the sets $\mathcal{A}$ through $\mathcal{E}$, and compare our results with previous simulations, in the frame of the DSA and SDA theories. Runs from the $\mathcal{F}$ and $\mathcal{G}$ sets were used in the direct comparison between PIC and hybrid results. Finally, extensive numerical convergence tests were also conducted for both codes. These tests included assessment of the effects of grid resolution, number of particles per cell, timestep, and numerical effects due to the finite box size (both in the longitudinal and transverse directions, as detailed in the Appendix).

%
%
%
%
\section{Global shock features}
\label{section3}
%

\begin{figure*}
\figurenum{1} 
\epsscale{1.}
\plotone{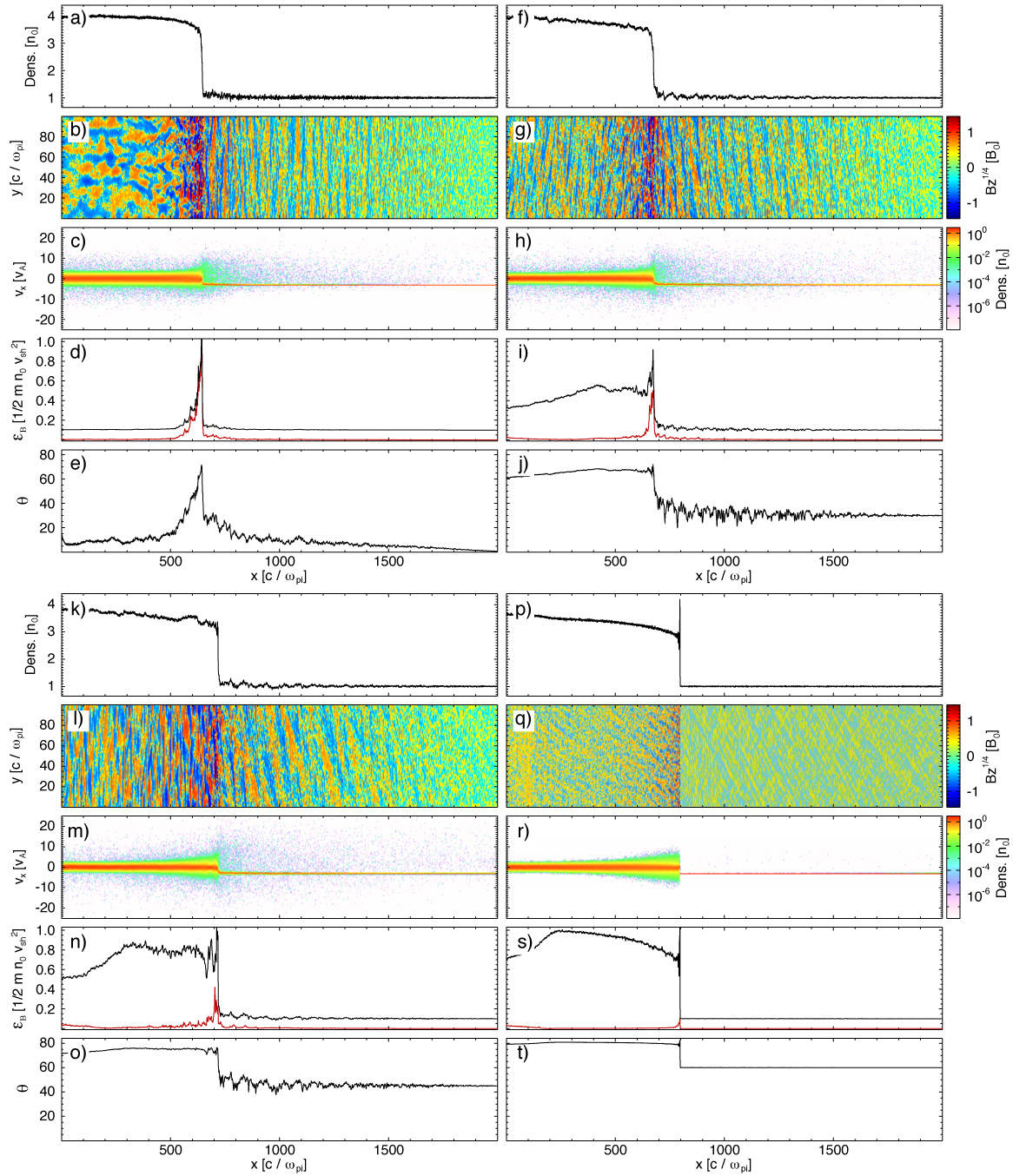}
\caption{Global shock features for 4 shocks, at time $t=600\,\mathrm{\omega_{ci}^{-1}}$, with $M_A=3.1$ and $\theta=0^\circ$ (run $\mathcal{A}_1$, upper left 5 panels), $30^\circ$ (run $\mathcal{A}_3$, upper right 5 panels),  $45^\circ$ (run $\mathcal{A}_4$, lower left 5 panels), and $60^\circ$ (run $\mathcal{C}_5$, lower right 5 panels). For each run, from top to bottom we show the plasma density averaged over the $y$ direction, $B_z$ magnetic field component, $x-v_x$ phase space, magnetic field energy normalized to the incoming flow energy (total in black, self-generated in red), and average magnetic field angle (i.e., the angle of the B field vector with the $x$ direction in each cell, $\theta=\arccos{[B_x/|B|]}$, averaged over the $y$ direction).\label{fig1}} 
\end{figure*}

Figure \ref{fig1} shows the transversely-averaged ion number density, $B_z$ magnetic field component, $x-v_x$ phase-space, magnetic field energy (normalized to the incoming flow energy), and the average magnetic field inclination angle (i.e., $\left<\arccos{[B_x/|B|]}\right>$) for Mach number $M_A=3.1$ shocks for different B field inclination angles ($\theta=0^\circ, 30^\circ, 45^\circ$ and $60^\circ$). The shock compression ratio is $n_d/n_u\sim 4$ in all cases, as expected from the MHD Rankine-Hugoniot relations\footnotemark. From the phase space plot it is clear that for $\theta=60^\circ$ there are essentially no particles being reflected at the shock (Figure \ref{fig1}r versus Figures \ref{fig1}c,h and m). The properties of the magnetic field are very different: waves and turbulence are observable for quasi-parallel shocks ($\theta \le 45^\circ$) in the pre-shock region upstream, and no magnetic field perturbations are seen for the quasi-perpendicular shock. These waves are generated by the interaction of two counter-streaming beams in the pre-shock region: the upstream plasma and the particles reflected from the shock front. The waves are the source for the magnetic field turbulence that is necessary for the DSA mechanism to work. Their propagation direction is always parallel to the background magnetic field, such that the wave fronts reach the shock surface at an angle for all $\theta>0^\circ$ (reading the propagation angle from Figure \ref{fig1} is non-trivial, since the simulation box is not plotted with correct aspect ratio).
\footnotetext{This ratio depends on the adiabatic index which for 2D magnetized case is the same as in 3D (particles' velocities are not confined to the simulation plane, and have 3 degrees of freedom in both cases).}

As the cold upstream plasma reaches the shock, it heats up and starts to thermalize to a higher temperature downstream. There is a temperature overshoot just after the shock jump, observable as a larger spread in the $x-v_x$ plots in Figures \ref{fig1}c,h,m and r, where $T_{t}>T_d$ (with $T_{t}$ and $T_d$ the temperatures in the transition and far downstream regions, respectively). The transition region where $T_{t}>T_d$ also corresponds to a zone of enhanced electromagnetic waves, which is observable as a sharp rise in the magnetic field energy in Figures \ref{fig1}d,i,n and s. Finally, Figures \ref{fig1}e,j,o and t also show the effective averaged B field angle along the shock direction. A gradual increase in effective $\theta$ is most evident for the $\theta=0^\circ$ run. This feature of quasi-parallel shocks is due to the waves that are generated in the upstream pre-shock region. The region of enhanced B field extends over a very large zone upstream in the $0^\circ$ case, and it significantly affects the incoming plasma. Although the far upstream B field angle varies up to $60^\circ$ for these runs, the mean B field angle at the shock boundary is nearly constant ($\sim70^\circ$). This then suggests that even strictly parallel shocks are dominated by perpendicular shock physics, near the shock front, due to the strong transverse field of self-generated circularly polarized upstream waves.
\begin{figure*} 
\figurenum{2} 
\epsscale{1.}
\plotone{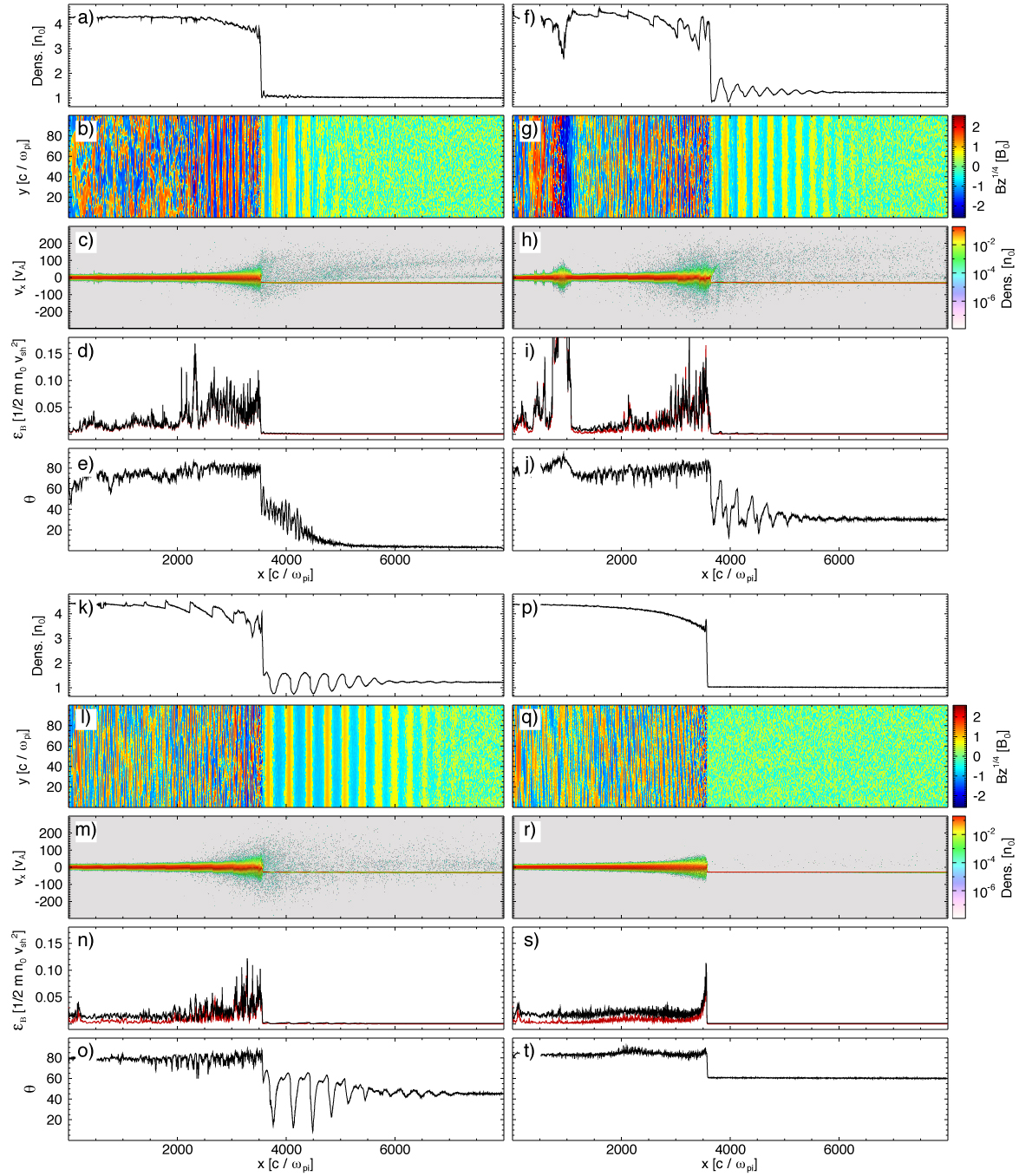}
\caption{Same as Figure \ref{fig1}, for shocks with $M_A=31$, and for the same inclination angles ($\theta=0^\circ$, $\theta=30^\circ$, $\theta=45^\circ$ and $\theta=60^\circ$), at time $t=360\,\mathrm{\omega_{ci}^{-1}}$. Waves in the upstream are affected by the limited transverse box size (see text).\label{fig1_1}} 
\end{figure*}

Figure \ref{fig1_1} shows the global shock structure for $M_A=31$ shocks (the frames are the same as for Figure \ref{fig1}). In this case the shock compression is still $\sim4$ as before, and the same suppression of reflected particles for angles $\theta > 45^\circ$ is observed. The most striking difference is in the self-generated magnetic field: the waves excited in the upstream reach higher intensities relative to the background magnetic field (this causes shock reformation in some cases, resulting in strong density fluctuations such as in Figure \ref{fig1_1}f-j at $x\sim1000\,\mathrm{c/\omega_{pi}}$). Also, the shock transition region is typically longer than for low Mach number shocks ($\sim100\,\mathrm{c/\omega_{pi}}=30\,\mathrm{r_{Li}}$ for $M_A=3.1$ in Figure \ref{fig1} compared to $\sim1500\,\mathrm{c/\omega_{pi}}=45\,\mathrm{r_{Li}}$ for $M_A=31$ in Figure \ref{fig1_1}). The magnetic field energy profile for $M_A=3.1$ and $M_A=31$ shocks (red lines in Figure \ref{fig1} and Figures \ref{fig1_1}d,i,n and s) is also quite different. For the higher Mach number shocks, the self-generated magnetic field energy (minus the component due to compression, red lines) peaks at $\sim10\%$ to $18\%$ of the incoming plasma flow energy, whereas this figure is $\sim50\%$ for the low Mach number case. Notice also that, although the $M_A=31$ shock simulations show the magnetic field propagating parallel to $x$, this is an artifact of the limited size in the y direction; we used larger boxes to confirm that waves propagate parallel to the background B field, and that no other significant differences are observed in the results (see Appendix).
\begin{figure} 
\figurenum{3} 
\epsscale{1.2}
\plotone{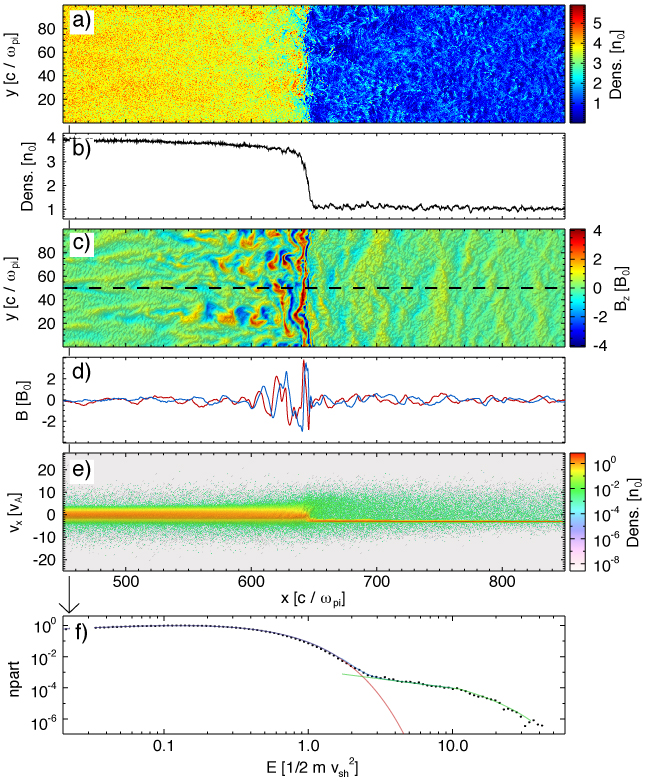}
\caption{Shock features for the $M_A=3.1$ parallel shock ($\theta=0^\circ$), run $\mathcal{A}_1$, at $t=600\,\mathrm{\omega_{ci}^{-1}}$. Panel a) shows the plasma density, panel b) shows the density averaged over the $y$ direction, panel c) shows the $B_z$ magnetic field component, panel d) shows the lineout of the $B_y$ (blue) and $B_z$ (red) magnetic field components (along the dashed line), and e) shows the $x-v_x$ phase space. Panel f) shows the particle spectra $\sim195\,\mathrm{c/\omega_{pi}}$ behind the shock (fitted to a Maxwellian, a power law, and an exponential cutoff).\label{fig2}} 
\end{figure}

\begin{figure} 
\figurenum{4} 
\epsscale{1.2}
\plotone{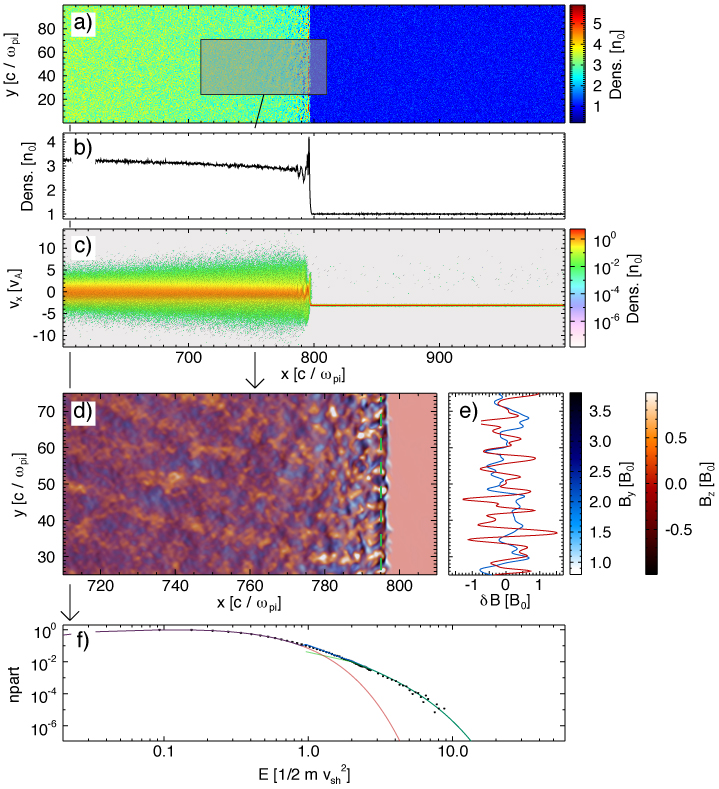}
\caption{General shock features for the $M_A=3.1$ quasi-perpendicular shock ($\theta=60^\circ$), run $\mathcal{A}_5$, at $t=600\,\mathrm{\omega_{ci}^{-1}}$. Panel a) shows the plasma density, panel b) shows the density averaged over the $y$ direction, and panel c) shows the $x-v_x$ phase space. Panel d) shows the $B_y$ (blue color table) and $B_z$ (red color table) magnetic field components in the zoomed-in box (as depicted in panel a). Panel e) shows the lineout of the two field components over the $y$ direction (along the green dashed line in panel d). Panel f) shows the particle spectra $\sim195\,\mathrm{c/\omega_{pi}}$ behind the shock (fitted to a Maxwellian, a power law, and an exponential cutoff).\label{fig3}} 
\end{figure}

Figures \ref{fig2} and \ref{fig3} show the detailed features for the quasi-parallel shock with $M_A=3.1$ and $\theta=0^\circ$, and for the quasi-perpendicular shock with the same Mach number and $\theta=60^\circ$. In the quasi-parallel case, panels c) and d) show a clear signature of right-hand circularly polarized Alfv\'en waves. In the simulation / downstream frame, the waves in the upstream are propagating toward the shock, parallel to the background magnetic field and with the B field vector rotating clockwise, when looking into the $-x$ direction (wave is resonant with the ions traveling in the $+x$ direction and $k=-k_x\,\vec{e}_x$, see Figure \ref{fig2}d). The effect of these waves on the plasma is visible as density perturbations in the upstream region (Figures \ref{fig2}a,b). The particle spectrum shown in Figure \ref{fig2}f (measured downstream at $x=450\,\mathrm{c/\omega_{pi}}$) is fitted to a Maxwellian component (in red), and a power-law component with an exponential cutoff (in green),
\begin{equation}
f(E)\propto E^{1/2} \mathrm{e}^{-\frac{E}{E_{th}}}+k_1 E^{-\alpha} \mathrm{e}^{-\frac{E}{E_{cut}}},\label{fit}
\end{equation}
where the power-law term is only added above an injection energy $E_{inj}$. The thermal spread for the Maxwellian distribution, expressed here and throughout the paper as a fraction of the incoming upstream plasma bulk-flow energy, is $E_{th}=0.39\,\mathrm{E_{up}}$ ($39\%$ of the upstream bulk energy). From Rankine-Hugoniot conditions the total energy spread in the ion distribution should be $\sim0.5\,\mathrm{E_{up}}$; from simulation results we instead find this value to be $\sim0.42\,\mathrm{E_{up}}$ (which includes the $0.39\,\mathrm{E_{up}}$ in the thermal component), with the remaining energy ($0.58\,\mathrm{E_{up}}$) being stored mostly in the electron component. The power-law index measured in this case is $\alpha\sim1.8$ and is not changing significantly for $t>200\,\mathrm{\omega_{ci}^{-1}}$. There are, however, both spatial and temporal fluctuations in the spectral slope of up to $\pm0.2$, if measurements are taken at different times  and / or different regions downstream. The general trend shows a slightly steeper spectral index $\alpha\sim2.2$ at earlier times, when the shock is forming. These variations are mostly due to numerical effects\footnotemark; the energy efficiency for a given shock, however, does not vary significantly when numerical convergence studies are performed.
\footnotetext{Varying numerical factors such as resolution and number of particles per cell alters the noise properties of hybrid codes, which in turn influences the thermalization of the ions downstream, and injection into the power-law.}

The strong contrast with quasi-perpendicular shocks is evident in Figure \ref{fig3}. The upstream plasma density (panel a) is not perturbed and EM turbulence upstream is non-existent (panel d); also there are no particles returning upstream (traveling in the $+x$ direction). Electromagnetic waves are seen only in the shock transition region, and are propagating parallel to the background B field in this zone (Figures \ref{fig3}d,e). The resulting particle spectra can be seen in Figure \ref{fig3}f. Again, there is a thermal spread $E_{th}=0.39\,\mathrm{E_{up}}$ in this case, which is the value expected from the shock jump conditions for this B field inclination angle. The non-thermal component of the spectrum, however, is very limited in range and the maximum particle energy growth in time is negligible; the green line in the plot is still a fit to a power-law with index $\alpha=2$, but with a very steep cutoff starting at $E_{cut}=2\,\mathrm{E_{up}}$, which indicates that indeed particle acceleration is almost non-existent in this case. Similar results are found for run $\mathcal{A}_6$ for the same Mach number and $\theta=75^\circ$.
%
%
%
%
%
\section{Shock parameter survey}
\label{section4}

The extensive parametric study conducted here includes shock Mach numbers from $M_A=3.1$ up to $M_A=100$, and both quasi-parallel and quasi-perpendicular shocks with angles from $0^\circ$ up to $75^\circ$, as specified in Table \ref{table1} (runs $\mathcal{A}$ through $\mathcal{E}$). The goal of this study is twofold: to understand how shock acceleration efficiency changes with these parameters and to understand the microphysics of particle acceleration for the most relevant scenarios. In order to understand the particle acceleration mechanisms, it is also crucial to understand how EM turbulence is generated, which relates closely to particle diffusion and acceleration in the DSA framework.
%
%
\subsection{Wave generation and shock reformation}

The dominant waves present in magnetized non-relativistic shocks are electromagnetic ion modes. There are two relevant modes to consider: the right-hand Alfv\'en mode, and the left-hand ion-whistler mode. The phase velocities for these modes in the hybrid limit are
\begin{eqnarray}
V_{RH}=V_A\sqrt{1-\frac{\omega}{\omega_{ci}}}\label{righthand}\\
V_{LH}=V_A\sqrt{1+\frac{\omega}{\omega_{ci}}}.\label{lefthand}
\end{eqnarray}
These can be derived from the full wave dispersion relation for a cold magnetized ion electron plasma, in the non-relativistic limit of $V_A\ll c$, considering a large ion to electron mass ratio, and typical frequencies $\omega<\omega_{ci}$. Both full PIC and hybrid simulations show similar wave structures, and exhibit the modes described by Eq. (\ref{righthand}-\ref{lefthand}); the agreement between the models is also very good (see Appendix). 
\begin{figure} 
\figurenum{5} 
\epsscale{1.2}
\plotone{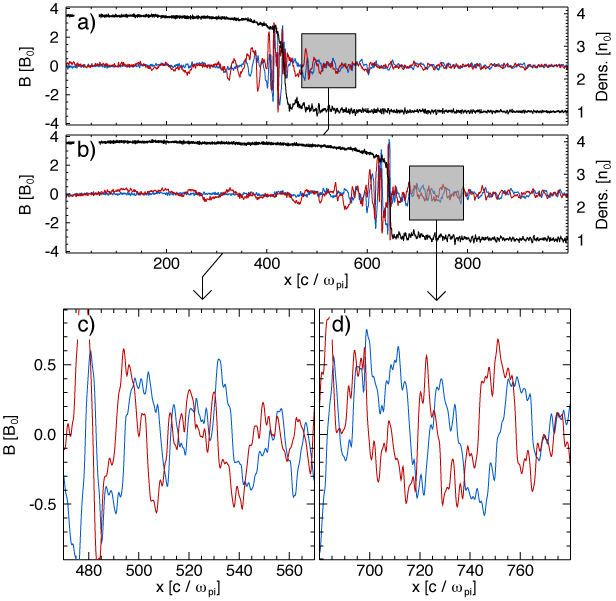}
\caption{Time evolution for the averaged density (black), $B_y$ (blue) and $B_z$ (red) magnetic field components at times $t=400\,\mathrm{\omega_{ci}^{-1}}$ (panel a), and $t=600\,\mathrm{\omega_{ci}^{-1}}$ (panel b) ($M_A=3.1$, $\theta=0^\circ$, run $\mathcal{A}_1$). Panels c) and d) show the magnetic field components at a fixed distance in front of the shock at the two times.\label{fig4}} 
\end{figure}

These wave modes are seen in Figure \ref{fig2} and Figure \ref{fig3}. As observed from the downstream simulation frame, the waves in the upstream are propagating towards the shock front along the background magnetic field; for the quasi-parallel $M_A=3.1$ shock case (Figure \ref{fig2}) the waves are right-hand circularly polarized Alfv\'en waves with $V_{ph}=V_{RH}$ from Eq. (\ref{righthand}) (i.e., when looking into the $\vec{k}$ direction of propagation at a fixed position in space, the $\vec{B}$ and $\vec{E}$ vectors turn clockwise in time).
\begin{figure} 
\figurenum{6} 
\epsscale{1.2}
\plotone{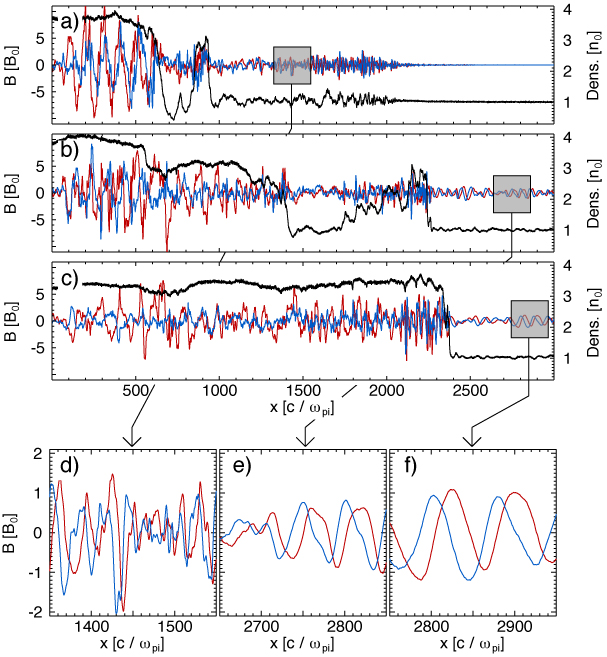}
\caption{Time evolution for the averaged density (black), $B_y$ (blue) and $B_z$ (red) magnetic field components at times $t=60\,\mathrm{\omega_{ci}^{-1}}$ (panel a), $t=120\,\mathrm{\omega_{ci}^{-1}}$ (panel b), and  $t=180\,\mathrm{\omega_{ci}^{-1}}$ (panel c)($M_A=31$, $\theta=0^\circ$, run $\mathcal{D}_1$). Panels d), e) and f) show the magnetic field components at a fixed distance from the shock front for the three time values.\label{fig5}} 
\end{figure}

Figure \ref{fig4} shows the averaged density and the lineout of the $B_y$ and $B_z$ magnetic field components at two different times for the same $\theta=0^\circ$, $M_A=3.1$ run. It is apparent that the shock structure is identical at $t=400\,\mathrm{\omega_{ci}^{-1}}$ and at $t=600\,\mathrm{\omega_{ci}^{-1}}$, and in fact this is true also for larger times up to the end of the simulation. Although the pre-shock region continues to extend farther into the upstream, the shock front propagates at a constant velocity and the shock evolution is continuous in time. The zoomed-in boxes in Figure \ref{fig4} also show that the largest wavelength in this upstream region (measured at a constant distance from the shock front) is $\lambda\sim30\,\mathrm{c/\omega_{pi}}$, and it is similar for the two times considered. These waves are resonant with the highest-energy particles (ions) propagating away from the shock (traveling in the $+x$ direction); this can be inferred from both Figure \ref{fig2} and Figure \ref{fig4}, by looking at the two components of the magnetic field. 

This behavior changes when we consider higher Mach number shocks. One to one comparison of the averaged density and magnetic field components for a $\theta=0^\circ$, $M_A=31$ shock can be seen in Figure \ref{fig5}, for three different instants in time. In this case, the waves in the upstream zoomed-in boxes are still propagating towards the shock, but instead they are left-hand circularly polarized ion-whistler waves with $V_{ph}=V_{LH}$ from Eq. (\ref{lefthand}), and they are non-resonant with the particles traveling away from the shock in the $+x$ direction. A similar behavior is seen for quasi-parallel shocks with $M_A=10$ and $M_A=100$. Also, shock evolution at later times in the high Mach number regime exhibits both the left-hand and the right-hand modes; which of the modes dominates depends on the local properties of the interacting flows, which vary with distance to the shock front - the resonant right-handed mode dominates on larger scales $\lambda\sim1000\,\mathrm{c/\omega_{pi}}$, while the non-resonant left-handed mode has a typical wavelength $\lambda\sim100\,\mathrm{c/\omega_{pi}}$.

One very important consequence of the different behavior between high and low Mach number shocks is visible in Figure \ref{fig5} for $M_A=31$ shock. The self-generated magnetic field amplitude is greater than the background magnetic field by $\delta B/B_0\sim2$ to $4$ in some zones upstream. This results in an effective magnetic field inclination angle $\theta>45^\circ$ in these zones, which in turn strongly affects the plasma flow. Figures \ref{fig5}a-c show strong density perturbations growing in the upstream pre-shock region, which eventually causes particles to reflect at these points and the shock front to reform. This discontinuous behavior is in strong contrast with the results for shocks with lower Mach number; a quasi-steady behavior of the shock structure in the case of high Mach numbers is not guaranteed, although it might be achieved at later times. Another indication that quasi-stationary behavior was not yet reached at $t=180\,\mathrm{\omega_{ci}^{-1}}$ is that the typical wavelengths observed in Figure \ref{fig5}d-f are growing in time from $\lambda\sim30\,\mathrm{c/\omega_{pi}}$ at $t=60\,\mathrm{\omega_{ci}^{-1}}$ up to $\lambda\sim75\,\mathrm{c/\omega_{pi}}$ at $t=180\,\mathrm{\omega_{ci}^{-1}}$. Notice, however, that the strong reformation process in Figure \ref{fig5} is a transient effect of the shock formation process, due to the initial low temperature of the two counter-propagating flows. Shock reformation later in time still occurs, but is more localized in space (shock reforms over distances $\sim100\,\mathrm{c/\omega_{pi}}$).

The strong non-resonant $\delta B/B_0$ magnetic field generation for high Mach numbers ($M_A>10$) resembles the Cosmic Ray Current Driven (CRCD) instability. This electromagnetic instability was first proposed by \citet{Bell:2004p24} in the context of magnetic field amplification in supernova remnant shocks, and further explored numerically in the non-linear phase by several authors \citep{reville,Stroman:2009p2260,Riquelme:2009p2089,Gargate:2010p1604}. Within the framework of this mechanism, the CRs non-resonantly excite an electromagnetic mode in the precursor of a SNR shock; the wavelengths are much smaller than the Larmor radius of the CR particles, and the polarization of the mode is the same as the ion-whistler wave branch described by Eq. (\ref{lefthand}). In our simulations, however, the typical wavelength measured in the precursor (which is growing in time as the shock evolves), is of the same order of magnitude as the Larmor radius of the lowest energy CR particles, and the waves propagate with a phase velocity close to $V_A$.
%

%
%
\subsection{Shock acceleration efficiency}

We define shock acceleration efficiency in two ways: as the fraction of energy in the tail of the particle distribution function to the total energy in the distribution (energy efficiency), and as the number of particles in the tail of the distribution to the total number of particles in the distribution (number efficiency). Spectra in the simulations can be fitted to Eq. (\ref{fit}), and the injection energy $E_{inj}$ is then used as a threshold to define which particles are part of the tail of the distribution, or part of the thermalized Maxwellian component. To assure consistency of measurements between different shocks, we measure the spectra by considering a shell of particles in the downstream region just beyond the shock transition region (shell width is $\sim2$ to $5\,\mathrm{r_{Li}}$, and extends up to $\sim100\,\mathrm{c/\omega_{pi}}$ from the shock front for $M_A=3.1$ shocks, and up to $\sim1500\,\mathrm{c/\omega_{pi}}$ for $M_A=31$ shocks).

The particle spectra are also measured at the same time $t=300\,\mathrm{\omega_{ci}^{-1}}$ across all shocks. This time was chosen due to physical and numerical considerations. On the one hand, before $t\sim40\,\mathrm{\omega_{ci}^{-1}}$ the shocks are still forming, and thus measuring particle spectra at this time is not interesting. On the other hand, due to the finite size of the simulation boxes in the $x$ direction, the production of energetic particles at the shock saturates after $t\sim600\,\mathrm{\omega_{ci}^{-1}}$ for our numerical parameters; this is a known effect due to the fact that the most energetic particles start to escape from the simulation box, which introduces a high-energy cutoff in the spectrum.
\begin{figure} 
\figurenum{7} 
\epsscale{1.2}
\plotone{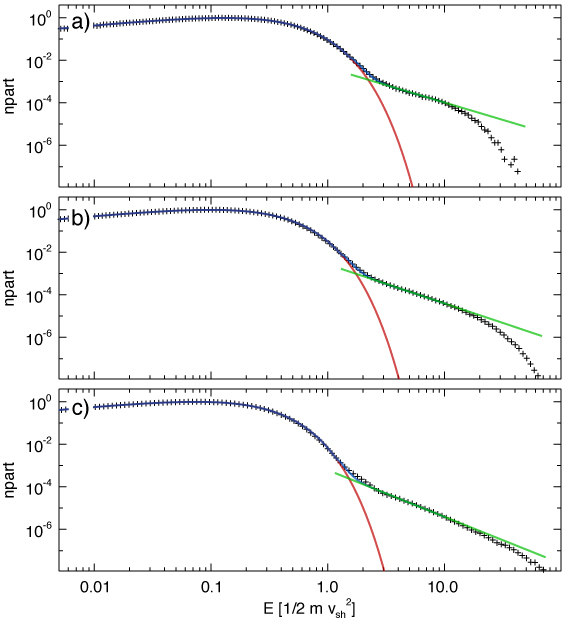}
\caption{Particle spectra (and function fits) measured in the downstream at time $t=300\,\mathrm{\omega_{ci}^{-1}}$ (black crosses). Spectra are shown for run $\mathcal{A}_1$ ($M_A=3.1$, $\theta=0^\circ$) in frame a), for run $\mathcal{C}_1$ ($M_A=10$, $\theta=0^\circ$) in frame b), and for run $\mathcal{D}_1$ ($M_A=31$, $\theta=0^\circ$) in frame c).\label{fig7}} 
\end{figure}
\begin{figure} 
\figurenum{8} 
\epsscale{1.2}
\plotone{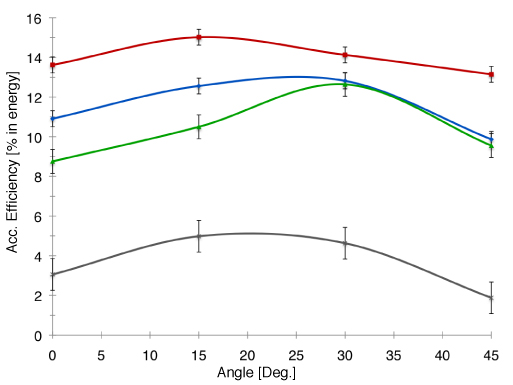}
\caption{Acceleration efficiency, defined as the fraction of energy in the power law tail to the total energy in the distribution, as a function of the shock magnetic field inclination angle. Efficiencies for runs $\mathcal{A}_1$ to $\mathcal{A}_4$ ($M_A=3.1$) are in blue, runs $\mathcal{B}_1$ to $\mathcal{B}_4$ ($M_A=6$) in red, runs $\mathcal{C}_1$ to $\mathcal{C}_4$ ($M_A=10$) in green, and runs $\mathcal{D}_1$ to $\mathcal{D}_4$ ($M_A=31$) in grey.\label{fig8}} 
\end{figure}

Figure \ref{fig7} compares particle spectra measured for parallel $\theta=0^\circ$ shocks with different Mach numbers, for  $M_A=3.1$ (frame a), $M_A=10$ (frame b), and $M_A=31$ (frame c), at time $t=300\,\mathrm{\omega_{ci}^{-1}}$. As time evolves, the maximum power-law tail energy grows and the temperature of the distribution decreases (Figure \ref{fig6}a in the Appendix shows this trend for a $M_A=4.7$ shock). The energy efficiencies derived from fits as the ones in Figure \ref{fig7} show a general decrease at larger Mach number; this is because there are fewer particles in the tail of the distribution (above the injection energy) as the Mach number increases.

The shock acceleration efficiency as a function of magnetic field inclination angle for the runs in sets $\mathcal{A}$ through $\mathcal{D}$ (Mach numbers from $3.1$ to $31$) can be seen in Figure \ref{fig8}. As a general trend, the acceleration efficiency decreases with increasing Mach number and with increasing angle. The highest energy efficiency measured was $E_{eff}\sim15\%$ for the $M_A=6$, $\theta=15^\circ$ shock. It is also apparent from Figure \ref{fig8} that there is a consistent increase in efficiency, for each Mach number, at an angle of $15^\circ$ to $30^\circ$. There is thus a critical intermediate angle for which the shock energy conversion efficiency is greatest. This can be understood in the framework of DSA and SDA theories: the SDA mechanism, which is faster than DSA, is more effective for higher B field inclination angles.

The energy efficiency decrease with shock Mach number is very clear from Figure \ref{fig8}. The trend is similar when looking at number-efficiency (fraction of particles in the tail to the total number of particles), which is seen in Figure \ref{fig7}, where the power law index is roughly the same $\alpha\sim2\pm0.2$ for all runs at $t=300\,\mathrm{\omega_{ci}^{-1}}$, but the number of particles in the tail decreases from frame a) ($M_A=3.1$) to frame c) ($M_A=31$). This result is an indication that particle injection into the DSA process is less efficient at higher Mach numbers due to the local conditions near the shock front. Analyzing the micro-structure of the shock transition, and understanding in detail the particle acceleration mechanism is thus of crucial importance. It should also be noted  that there is a a slight overall increase in the efficiency from $M_A=3.1$ to $M_A=6$ shocks (Figure \ref{fig8} blue line to the red line) which is contrary to the general trend; a similar increase is found in \citet{GIACALONE:1997p1192} at these Mach numbers. This is due to two effects: on the one hand there is always an increase of the maximum energy in the ion distribution function as the Mach number increases, but on the other hand there is a consistent decrease of the number of accelerated particles in the tail for higher Mach numbers. Since $\alpha\sim2$, there is an equal amount of energy per decade in the distribution and thus high energy particles contribute significantly to the energy efficiency. The two effects are then both relevant. At low Mach numbers from $M_A=3.1$ to $M_A=6$, the first effect dominates, whereas for $M_A=10$ and beyond the second effect dominates, which then results in the overall efficiency decrease seen in Figure \ref{fig8}.

For higher Mach number shocks ($M_A=100$), the general shock features, such as the electromagnetic field structure, including the polarization of the waves in the upstream and the shock compression ratio are consistent with the shock jump conditions (and are similar to $M_A=31$ shocks). However, in this case, we found that shock evolution would have to be followed for longer times and using a longer simulation box for a consistent measurement of the efficiency; tentative values in the order of  $\sim1\%$ were found at the end of the simulations at $t=80\,\mathrm{\omega_{ci}^{-1}}$ for $M_A=100$ shocks. For shocks with angles $\theta>45^\circ$ we found that the power law was strongly suppressed.  
%
%
%
%
\section{Particle acceleration}
\label{section5}

As the particle acceleration and the shock evolution are dynamic processes, it is expected that the maximum particle energy in the power-law tail should grow in time. For a DSA-accelerated particle, the total elapsed time for a particle to accelerate from momentum $p_0$ to $p$ depends on the diffusion coefficient and can be estimated as \citep{DRURY:1983p2322}
\begin{equation}
\tau_{acc}(p)=\frac{3}{\Delta V}\int_{p_0}^{p}{\left(\frac{\kappa_u(p^{'})}{V_u}+\frac{\kappa_d(p^{'})}{V_d}\right)\frac{d p^{'}}{p^{'}}}
\label{drury}
\end{equation}
where $V_u$, $V_d$ are the upstream and downstream flow velocities in the shock frame, and $\kappa_u$, $\kappa_d$ are the upstream and downstream spatial diffusion coefficients along the shock propagation direction. If a Bohm-type diffusion coefficient is now considered, $\kappa=1/3 \lambda v$, where $\lambda=r_{Li}$ is the particle's mean free path and $v$ the particle's velocity, it is possible to deduce $\tau_{acc}\propto E$; that is, under Bohm diffusion, the energy of a particle undergoing diffusive shock acceleration will grow linearly with time.
\begin{figure} 
\figurenum{9} 
\epsscale{1.2}
\plotone{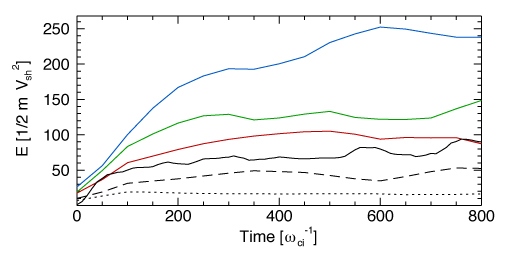}
\caption{Evolution of maximum particle energy in time, for shocks with $M_A=3.1$ (runs from set $\mathcal{A}$), for angles $\theta=0^\circ$ (solid black line), $\theta=15^\circ$ (red), $\theta=30^\circ$ (green), $\theta=45^\circ$ (blue), $\theta=60^\circ$ (dashed line), and $\theta=75^\circ$ (dotted line).\label{fig9}} 
\end{figure}

Figure \ref{fig9} shows the evolution of the energy of the most energetic particle in a given simulation for shocks with different B field inclination angles. Energy evolution in time is fastest for the $M_A=3.1$, $45^\circ$ shock, which also corresponds to the case where the largest energy is attained. This is partly due to the SDA mechanism: the upstream background electric field intensity will be higher for higher inclination angles. As it can be seen in Figure \ref{fig8} (blue line), however, the $M_A=3.1$, $45^\circ$ shock is not the one with the best acceleration efficiency; this is due to the fact that injection into DSA mechanism is harder for higher inclination angles, such that a lower number of particles are present in the power-law tail. The evolution of energy in time for the $M_A=3.1$ shocks with intermediate inclination angles $\theta\sim15^\circ$ to $30^\circ$ is slower than for the same shock with $\theta=45^\circ$, and faster than for lower angles. Since injection into the DSA mechanism is more effective at lower angles, the end result is that shocks with angles in the range of $\sim15^\circ$ to $30^\circ$ are the most efficient particle accelerators, for the parameters surveyed.
\begin{figure} 
\figurenum{10} 
\epsscale{1.2}
\plotone{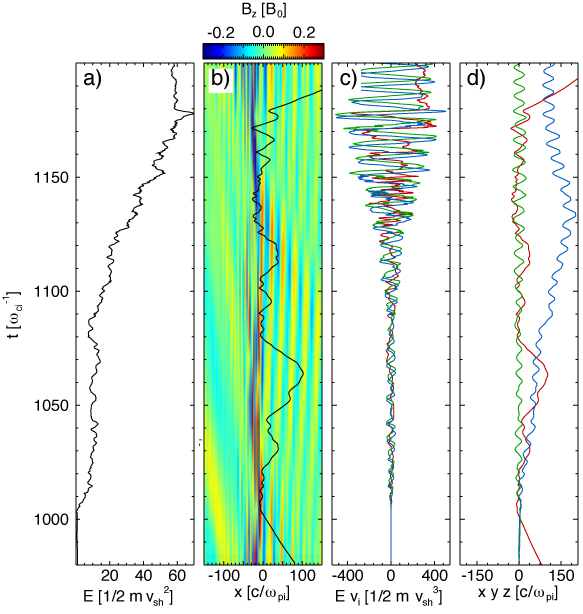}
\caption{Time evolution of key characteristics for a typical accelerated particle, extracted from run $\mathcal{A}_1$ ($M_A=3.1$, $\theta=0^\circ$). From left to right we show: a) particle energy, b) particle distance from the shock (in the shock propagation direction), c) the product of the Energy and the particle's velocity components $E\,v_x$ (red), $E v_y$ (green), and $E v_z$ (blue), and d) the displacement of the particle in the $x$ (red), in the $y$ (green) and the $z$ (blue) direction. Panel b) also shows the $B_z$ magnetic field component time evolution at the shock front as color.\label{fig10}} 
\end{figure}

By measuring the maximum energy in the particle distribution at several points in time in one simulation, it is possible to estimate what is the energy gain in time. For the runs with $M_A=3.1$ we find that the energy evolution follows a power-law in time, as $E\propto t^\beta$, except for angles $\theta\ge 60$, which show almost no energy growth in time. For the $M_A=3.1$, $\theta=0^\circ$ run, we get $\beta\sim0.28$, which implies that diffusion is less efficient than Bohm (i.e., using Eq. (\ref{drury}), $\beta< \beta_{B}=1$), in the approximation that all the energy gain is due to DSA. This measurement yields similar results for other B field inclination angles; a systematic study of the variation of $\beta$, and of the diffusion coefficient itself, with Mach number and inclination angle will be the subject of future work, since this study would imply a different and more thorough analysis. 
\begin{figure} 
\figurenum{11} 
\epsscale{1.2}
\plotone{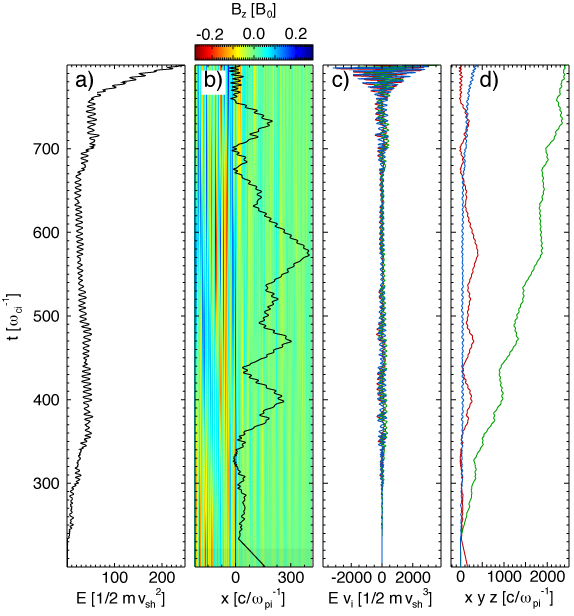}
\caption{Time evolution of key characteristics for a typical accelerated particle, extracted from run $\mathcal{A}_4$ ($M_A=3.1$, $\theta=45^\circ$). Quantities plotted are the same as in Figure \ref{fig10}.\label{fig11}} 
\end{figure}

The trajectories of energetic simulation particles support the claim that the DSA mechanism is responsible for particle acceleration in non-relativistic magnetized shocks. Particle track information is presented in Figures \ref{fig10} and \ref{fig11}, for typical high-energy particles in $M_A=3.1$ shocks with $\theta=0^\circ$ and $\theta=45^\circ$, respectively. These resemble the particle tracks and acceleration mechanism described in \citet{Sugiyama:2011p3494}, where particles are greatly accelerated only within an effective distance $\Delta x_{eff}$ from the shock front.

Figure \ref{fig10}a shows the energy increase with time, from around the injection time at $t\sim1000\,\mathrm{\omega_{ci}^{-1}}$, which corresponds to the first crossing of the shock, until the particle escapes into the upstream medium. Multiple upstream/downstream reflections can be observed in Figure \ref{fig10}b, confirming a diffusive-type acceleration process. Most of the energy gain for this representative particle is achieved within a distance $\Delta x_{eff}\sim30\,\mathrm{c/\omega_{pi}}$ of the shock front, where the electromagnetic waves are more intense, and thus are more effective in scattering particles back to the shock. Also, this energy gain was achieved mostly in the time interval from $t\sim1090\,\mathrm{\omega_{ci}^{-1}}$ up to $t\sim1180\,\mathrm{\omega_{ci}^{-1}}$, and was approximately linearly proportional to time. For this particular shock case, where $\theta=0^\circ$ far upstream, particle confinement near the shock front is greatly enhanced by the self-generated EM waves which effectively increase the averaged magnetic field inclination angle in a region $\Delta x_{up}\sim100\,\mathrm{c/\omega_{pi}}$ upstream of the shock front (see Figure \ref{fig1}e). Thus, although the energy gain for individual particles is approximately linearly proportional to time over finite time intervals, the overall energy gain for the ensemble of accelerated particles over longer time intervals follows $E\propto t^{0.28}$ because, most of the time, the particles are not in the $\Delta x_{eff}$ upstream region where acceleration is effective.

In the case of the $M_A=3.1$, $\theta=45^\circ$ shock, Figure \ref{fig11}, accelerated particles also undergo several reflections upstream, and are carried back to the shock. In this case, the particles are effectively reflected back to the shock up to larger distances $\Delta x_{up}\sim300\,\mathrm{c/\omega_{pi}}$. Since the power and polarization of the EM waves generated due to the shock is similar between the $\theta=45^\circ$ and $\theta=0^\circ$ cases, the increase in the characteristic length for effective reflection is just due to the higher B field inclination angle (i.e., particles are more likely to be carried back to the shock when $\theta$ is higher). 

Overall, the global physical picture of the dependence of ion acceleration on the magnetic field angle is as follows. The upstream cold flow streaming towards the shock will scatter at the shock front, and there will be fewer particles escaping back upstream with increasing angle. This greater injection efficiency for lower angles is theoretically described in the literature \citep[e.g.,][]{GIACALONE:1996p2199}: for a given velocity and B field intensity, a larger angle implies a larger B field component parallel to the shock front, which will prevent an increasing number of particles from streaming away from the shock in the upstream direction. Once a particle is injected into the acceleration mechanism, however, a larger $\theta$ will keep higher energy particles closer to the shock front where they will be accelerated faster. Our results show just that: the typical acceleration rate for a $\theta=0^\circ$ case is $\Delta E/\Delta t\sim0.6\,\mathrm{E_{up}\,\omega_{ci}}$ (Figure \ref{fig10} from $t\sim1090$ to $t\sim1180,\mathrm{\omega_{ci}^{-1}}$); for a $\theta=45^\circ$ shock this rate is  $\Delta E/\Delta t\sim5\,\mathrm{E_{up}\,\omega_{ci}}$ (Figure \ref{fig11} for $t>750\,\mathrm{\omega_{ci}^{-1}}$). The acceleration of individual particles is due to the upstream convective electric field (which exists even for $\theta=0$ due to upstream waves), and thus the energy gain is directly proportional to the electric field strength and to time. The overall particle acceleration is slower (i.e.,$E\propto\,t^{\beta}$, with $\beta\sim0.28$), and is only slightly faster for larger angles due to the increasing intensity of the convective electric field.

Particle injection into the acceleration mechanism also changes with the shock Mach number. By analyzing a large number of trajectories, for both high and low Mach numbers at angles $\theta \le 45^\circ$, it is apparent that almost all particles are injected into the acceleration mechanism by reflection at the shock front, instead of thermal leakage from the downstream side. This thermal leakage process, which depends on the micro-physics of the shock structure, is a parameter in Diffusion Shock Acceleration theory\citep{Malkov:2001p2452,Amato:2005p2587,Amato:2006p2595}, and is assumed not to vary with shock Mach number. Our simulations, however, show that the number of particles that are actually injected into the DSA mechanism decreases with Mach number, so that the number efficiency varies from $N_{eff}\sim4.5\%$ for $M_A=3.1$, $\theta=0^\circ$ down to $N_{eff}\sim1\%$ for $M_A=31$, $\theta=0^\circ$, and even lower $N_{eff}\sim0.12\%$ for $M_A=100$, $\theta=0^\circ$ although measured at an earlier time. 

The physical reason for this decrease in injection efficiency with Mach number can be understood by analyzing the magnetic field energy at the shock front. In the shock transition region this energy, expressed as a fraction of the incoming flow energy, will increase relative to the far upstream value, due to the waves present in this zone (see for example, the B field profile in the shock transition region in Figures \ref{fig4} and \ref{fig5}, or the 2D profile of the same waves in Figures \ref{fig1} and \ref{fig2}). However, the average fraction of energy in these waves, relative to the upstream flow energy, decreases from $\Delta E_{wave}/ E_{up}\sim50\%$ for the $M_A=3.1$, $\theta=0^\circ$ case, down to $\Delta E_{wave}/ E_{up}\sim10\%$ for the $M_A=31$, $\theta=0^\circ$ case, and even lower $\Delta E_{wave}/ E_{up}\sim2.5\%$ for $M_A=100$, $\theta=0^\circ$. Also, for lower Mach numbers a sharper magnetic barrier is observed at the shock, while for higher Mach numbers the lower $\Delta E_{wave}/ E_{up}$ magnetic barrier extends further downstream. Or, stated in a different way, the shock transition region is longer (measured in ion skin depths or as a fraction of the upstream Larmor radius) for higher Mach number shocks. Typical cold upstream particles are then more easily specularly-reflected for low Mach numbers than for higher Mach numbers, which results in a larger fraction of particles being injected into the DSA mechanism for low Mach numbers. 
%
%
%
\section{Discussion and conclusions}
\label{section6}

In this paper we have discussed ion acceleration in non-relativistic magnetized shocks. In an effort to understand particle acceleration, we conducted a complete 2D survey of this class of shocks, varying both the Alfv\'enic Mach number and the magnetic field inclination angle for each shock. As shown before by other authors, our results confirm that there are fundamental differences between quasi-parallel and quasi-perpendicular shocks. Our study allows to systematically trace these differences as functions of the flow parameters and see the transition between different regimes. For quasi-parallel shocks, a significant number of particles are reflected by the shock back into the upstream medium, which causes electromagnetic instabilities that propagate back to the shock, and create a turbulent upstream medium. In contrast, for quasi-perpendicular shocks, with $\theta \ge 60^\circ$, there is no significant particle population being reflected at the shock front, and hence EM turbulence upstream is greatly reduced. 

Our results show a fundamental difference between shocks at low $M_A \le 10$ Mach numbers, and shocks with $M_A > 10$; for low Mach number shocks a right-hand mode with $\delta B/B_0<1$ is excited in the upstream, resonant with the ions streaming away from the shock, whereas for higher Mach numbers a left-hand non-resonant mode predominates. In the high Mach number case, the left-hand non-resonant waves exhibit B field amplitudes $\delta B/B_0>1$; these waves resemble the CRCD instability, as they have the same polarization. Here however, the wavelength $\lambda$ is comparable to the Larmor radius of the lowest-energy CR particles, and the phase velocity of the waves is around the Alfv\'en velocity. Thus, from the point of view of the highest-energy CR particles, since $\delta B/B_0>1$, $\lambda<r_{Li}$ and $V_A \ll V_{flow}$ (i.e.,the phase velocity of the waves is much less than the upstream flow velocity), the ion-whistler waves observed should have similar scattering properties to CRCD-generated electromagnetic turbulence.

For high Mach number shocks the self generated EM waves with $\delta B/B_0>1$ in the upstream medium start to reflect particles locally, and the shock is observed to reform further upstream; a similar effect of shock reformation due to Short Large-Amplitude Magnetic Structures (SLAMS) has also been described in the literature \citep{Dubouloz:1995p1593}. This shock jump is observed at a relatively early stage $t\sim100\,\mathrm{\omega_{ci}^{-1}}$ of the shock propagation, and it is thus possible that this is a transient effect due to the shock formation process. There are indications, from longer simulations (in time) over larger spatial domains that this is the case: although reformation still happens at late times, its effects are less noticeable than at the time the shock is forming. These simulations also show that the dominant EM modes change in time if the particle flux changes as well, which leads to a highly dynamic shock structure where sub-shocks form in localized regions of the upstream medium. Thorough analysis of such effects will be the subject of a future publication.

We compared shock acceleration efficiencies in similar conditions for all Mach numbers and angles, as described in Table \ref{table1}, and showed that the highest energy-conversion efficiencies are achieved for intermediate angles $\theta\sim15^\circ$ to $30^\circ$ and low Mach numbers. The power law indexes are similar in most cases, yielding $\alpha=2\pm0.2$. The typical growth of maximum energy in time is found to be $E\propto t^{0.28}$ (for $M_A=3.1$ and $\theta=0^\circ$). This value is indicative of a diffusion process less effective than Bohm diffusion, which would correspond to $E\propto t$; a full comparison of diffusion coefficients for different Mach numbers and inclination angles will also be the subject of future work.

The analysis of representative particle tracks for two quasi-parallel low Mach number shocks is consistent with DSA mechanism, and confirms that particles are reflected multiple times towards the shock front. Overall, we can conclude that ion acceleration in magnetized non-relativistic shocks is efficient at finite intermediate angles, in the quasi-parallel regime, and for low Mach numbers ($M_A<10$ ). This is explainable if we take into account that injection into a diffusive acceleration process is more likely at low angles, since a significant number of particles are allowed to travel farther into the upstream medium, and generate EM waves and turbulence which enhance further particle reflection towards the shock. For higher inclination angles, however, a particle with a sufficiently high energy will be more easily confined to the shock front, and will be able to accelerate faster due to stronger shock drifting (Figure \ref{fig9}); in terms of energy-conversion efficiency an intermediate angle yields the best results. Also, although higher Mach number shocks are able to accelerate particles to higher energies (as a fraction of incoming beam energy), the shock efficiency is lower because injection into the acceleration mechanism is less efficient at higher Mach numbers.

It is interesting to compare the present results with a similar shock survey done in \citet{GIACALONE:1997p1192} (henceforth G97) using 1D hybrid simulations for typical solar wind flow parameters. For a more limited range of Mach numbers, from $M_A=1.5$ up to $M_A=12$ the authors also show downstream spectra of accelerated particles; the power-law tails extend over a similar range in energy as in our case, although high-energy particle splitting, and the introduction of external EM turbulence was used to facilitate the diffusive process in the case of G97. In the comparable range of Mach numbers (up to $M_A=12$), the conclusions about shock acceleration efficiency are similar. At higher Mach numbers (not explored in G97) we observe a clear decrease in efficiency, and different features such as shock reformation and the left-hand wave polarization mode.

Energetic arguments, in order to explain CR energies of up to $10^{15}\,\mathrm{eV}$ from SNR sources, usually require a large fraction $> 30 - 40\%$ of the total energy in the shock to be transferred to accelerated particles, which indicates that shocks with $M_A < 30$ and $\theta<30^\circ$ are the most likely candidates for CR acceleration. For angles greater than $45^\circ$, injection is very limited and no significant acceleration is observed; it should be noted, however, that if pre-accelerated particles could be injected in shocks with $\theta\sim60^\circ$ in sufficient numbers, it would be possible to have higher acceleration efficiencies due to enhanced shock drift acceleration in the shock front (although this does not seem to occur naturally in the simulations). 

Relatively low Mach numbers ($M_A < 30$) are thus required for effective ion acceleration, which then implies that strong magnetic field amplification should occur at SNR shock sites (if we consider typical outflow velocities of $\sim1000\,\mathrm{km/s}$ in interstellar fields of $1\,\mathrm{\mu G}$). This strong B field amplification is inferred from observations \citep[for a review see][]{Reynolds:2011p3516}, and it might be achievable by the CRCD instability \citep{Bell:2004p24,reville,Stroman:2009p2260,Riquelme:2009p2089,Gargate:2010p1604}. Such low Mach numbers are also favored by the electron acceleration mechanisms in shocks \citep{Riquelme:2011p3505}. We do see strong B field amplification in our current simulations, but the effective Mach number near the shock front is at most reduced by half (effective Mach number near the shock is $M_A\sim15$, corresponding $\delta B\sim2 B_0$, for the case of $M_A=31$, $\theta=0$ far upstream conditions). A magnetic field amplification of at least an order of magnitude would be necessary in a SNR shock. As such, a self-consistent shock simulation where both accelerated particles and the CRCD instability are observed and where $\delta B\sim10 B_0$ is still lacking.

The numerical properties of the hybrid and PIC codes must be very well understood in order to validate results. We show the effects of dimensionality, and validate our choice of numerical parameters in the Appendix. As future work it would also be interesting to compare the current results with 3D simulations, in particular regarding the diffusion properties which can be different in 3D \citep{Jokipii:1993p3406}, and the characteristics of the wave spectra generated for quasi-parallel shocks, both in the low and high Mach number regimes. Finally, a complete statistical study of self-consistently accelerated particles in hybrid and PIC simulations would be important in order to assess the dependence of the diffusion coefficient on shock properties (such as the Mach number and B field inclination angle), and also on the individual particle energies. Such a study could then be used to identify the astrophysical relevant scenarios for the acceleration of cosmic rays.    
%
%

%
%

\section{Acknowledgments}
The authors would like to thank L. Sironi and D. Giannios for insightful comments and useful discussions, and R. Fonseca for the use of the visXD visualization routines \citep{Al:2008p3514}. This work was supported by NSF grant AST-0807381. The simulations presented in this article were performed on computational resources supported by the PICSciE-OIT High Performance Computing Center and Visualization Laboratory, the National Energy Research Scientific Computing Center, which is supported by the Office of Science of the US Department of Energy under contract No. DE-AC02-05CH11231, and Teragrid's Kraken under contract No TG-AST100035. 

\appendix

\section{Comparison of PIC and hybrid results}
\label{section7}

Comparing two different simulation models is useful to determine the domain of validity of the models. This kind of analysis has been done in the context of shocks for 1D hybrid and Monte Carlo simulations in \citet{ELLISON:1993p40}, and for 2D hybrid and PIC simulations of strictly perpendicular shocks in \citet{Hellinger:2007p2167}. Here we focus on comparing the hybrid and fully-kinetic methods for strictly parallel shocks, and in particular on how the shock structure and measured particle spectra is affected by the models and the choice of numerical parameters. 
\begin{figure} 
\figurenum{12} 
\epsscale{1.15}
\plotone{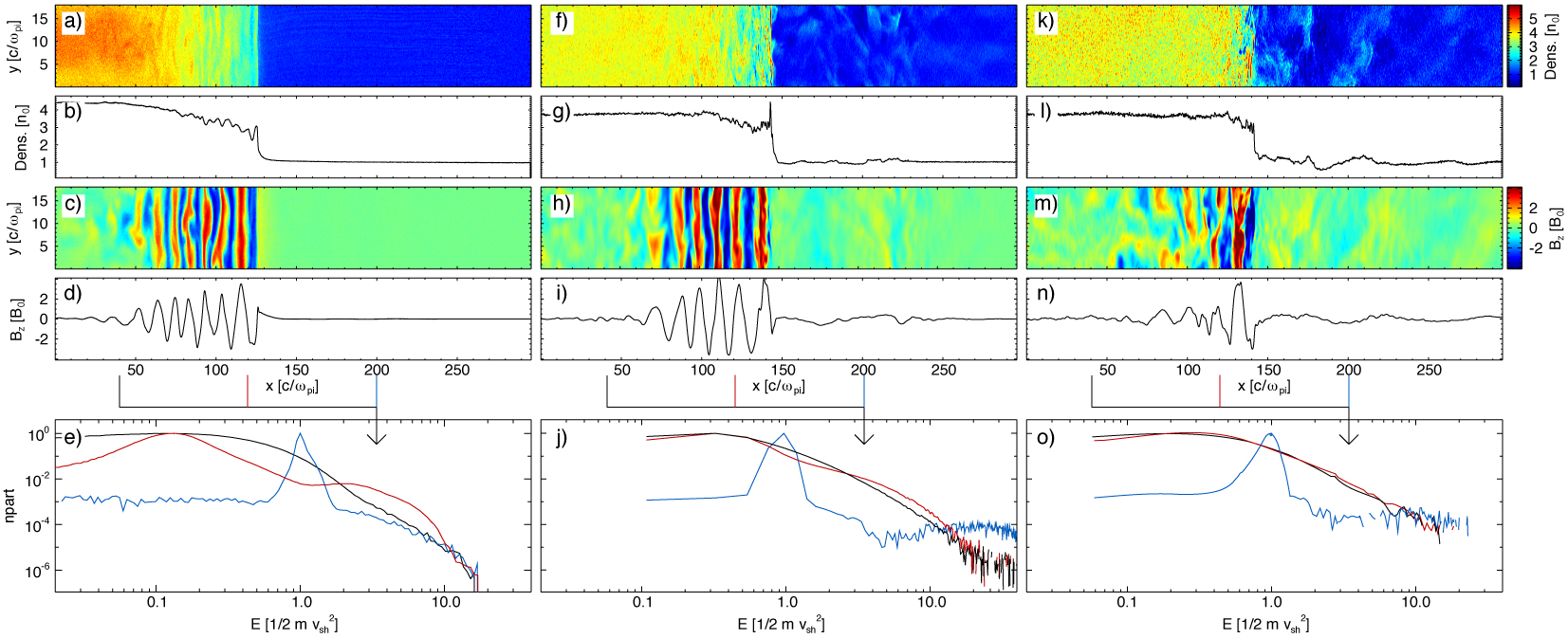}
\caption{Global shock features for a parallel ($\theta=0^\circ$) shock with Mach number $M_A=4.7$ at $t\sim87\mathrm{\omega_{ci}^{-1}}$ for a \textit{Tristan-MP} fully kinetic run ($\mathcal{F}$) (left panels), an hybrid run ($\mathcal{G}_1$) with the same numerical parameters (middle panels), and an hybrid run ($\mathcal{G}_2$) with an extended box size in the $x$ direction. From top to bottom the frames show: the plasma density, the plasma density averaged over the transverse direction $y$, the $B_z$ component of the magnetic field, $B_z$ averaged over the $y$ direction, and the particle spectra at three different positions ($x=40\mathrm{c/\omega_{pi}}$ in black, $x=120\mathrm{c/\omega_{pi}}$ in red, and $x=200\mathrm{c/\omega_{pi}}$ in blue).\label{fig12}} 
\end{figure}

We chose to compare a  $M_A=4.7$, $\theta=0^\circ$ shock. The agreement between the two models regarding the general features of the shock is very good, as it would be expected, for similar numerical parameters. The shock compression ratio is $n_d/n_u\sim4$, propagating at $V\sim1.6\,\mathrm{V_A}$ in the downstream reference frame; also in agreement with the Rankine-Hugoniot conditions for a parallel magnetized shock. Figure \ref{fig12} shows the ion density and $B_z$ magnetic field component for runs $\mathcal{F}$, $\mathcal{G}_1$ and $\mathcal{G}_2$ (from left to right, the top 4 panels), with all quantities plotted using the same scales. For runs $\mathcal{A}$ and $\mathcal{B}_1$ the numerical parameters are the same (namely, the same box size $296\,\mathrm{c/\omega_{pi}}\times18\,\mathrm{c/\omega_{pi}}$ is used), and both the shock structure and the waves in the shock transition region are very similar. However, when we consider a larger $L_x=2000\,\mathrm{c/\omega_{pi}}$ box size for run $\mathcal{G}_2$ (right panels in Fig \ref{fig12}), the size of the shock transition region is reduced significantly. The difference in the wave structure in the shock transition region, in this case, is due to the fact that the pre-shock region is not fully resolved for the smaller box size. This zone is strongly affected by the particles reflected at the shock front and traveling in the upstream medium towards $+x$; by limiting the box size to $L_x=296\,\mathrm{c/\omega_{pi}}$ in $\mathcal{F}$ and $\mathcal{G}_1$, the pre-shock zone is also limited. This limitation in turn affects the incoming particles (in the upstream, traveling towards the shock), which in the realistic case of run $\mathcal{G}_2$ start to gyrate before reaching the shock, and thus forming a less coherent beam (which leads to less coherent waves).
\begin{figure} 
\figurenum{13} 
\epsscale{.6}
\plotone{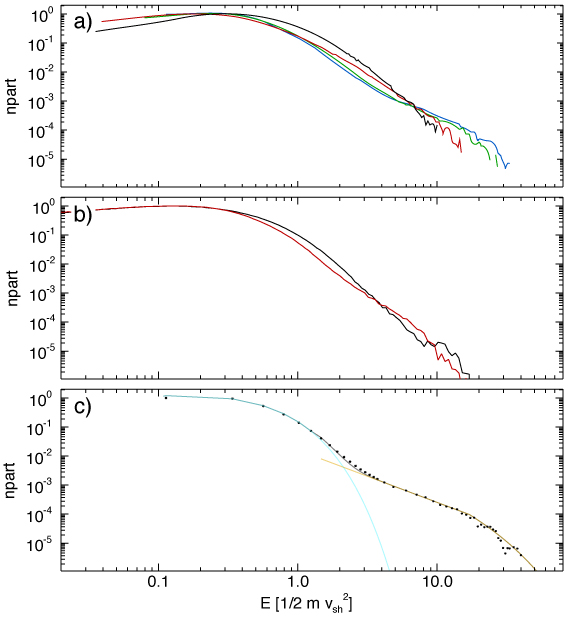}
\caption{Ion spectra for a $M_A=4.7$, $\theta=0^\circ$ shock, at times $t=40\,\mathrm{\omega_{ci}^{-1}}$ (black), $t=87\,\mathrm{\omega_{ci}^{-1}}$ (red), $t=200\,\mathrm{\omega_{ci}^{-1}}$ (green), and $t=400\,\mathrm{\omega_{ci}^{-1}}$ (blue) for hybrid run $\mathcal{G}_3$ (panel a). Panel b) shows the spectra at the same times ($t=40\,\mathrm{\omega_{ci}^{-1}}$ in black and $t=87\,\mathrm{\omega_{ci}^{-1}}$ in red) for the PIC run $\mathcal{F}$. Panel c) shows the hybrid spectra from run $\mathcal{G}_3$ at time $t=400\,\mathrm{\omega_{ci}^{-1}}$ fitted with a Maxwellian, a power law, and an exponential cutoff.\label{fig6}} 
\end{figure}

We show the measured particle spectrum at $x=40\,\mathrm{c/\omega_{pi}}$ in the downstream, $x=120\,\mathrm{c/\omega_{pi}}$ in the transition region and $x=200\,\mathrm{c/\omega_{pi}}$ in the upstream for the three runs in the bottom panels of Figure \ref{fig12}. The small differences between the spectra measured in the three runs $\mathcal{A}$, $\mathcal{B}_1$ and $\mathcal{B}_2$ are due to the differing number of particles used in plotting the spectra (less particles are being used to plot the spectra in the hybrid case). The downstream spectra (black lines in the bottom frames of Figure \ref{fig2}) can be modeled by a Maxwellian with a thermal spread of $E_{th}\sim0.3\,\mathrm{E_{up}}$ ($\sim30\%$ of the incoming flow bulk energy), a power law tail $f(E)\propto E^{-\alpha}$ and an exponential upper energy cutoff.      
\begin{figure} 
\figurenum{14} 
\epsscale{0.6}
\plotone{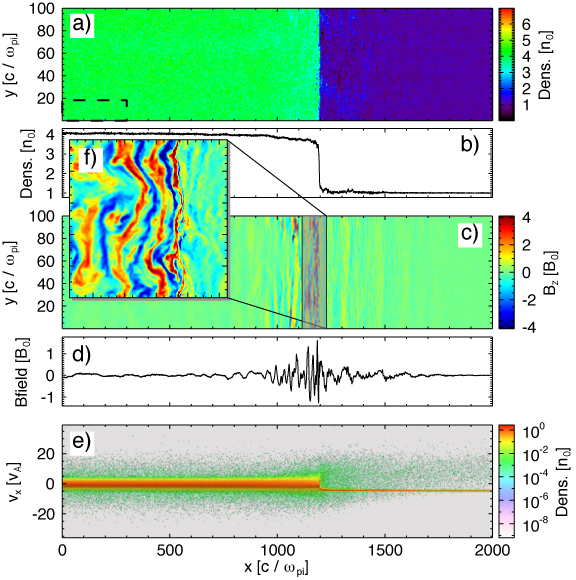}
\caption{Global shock features for an hybrid parallel ($\theta=0^\circ$) shock with Mach number $M_A=4.7$ at $t\sim\mathrm{\omega_{ci}^{-1}}$ (run $\mathcal{G}_3$). Numerical parameters are the same as in run $\mathcal{G}_2$, but an extended $y$ dimension is used. From top to bottom, the panels show: a) plasma density, b) plasma density averaged over $y$, c) $B_z$ component of the magnetic field,  d) $B_z$ averaged over the $y$ direction, and e) $x-v_x$ phase space. The zoomed in frame f) shows $B_z$ in a $100\times100\,\mathrm{(c/\omega_{pi})^2}$ box at the shock front.\label{fig13}} 
\end{figure}

The time evolution of the downstream particle spectrum can be seen in Figure \ref{fig6}. Panel a) corresponds to the hybrid run $\mathcal{G}_3$, which reproduces the physical parameters of the PIC run $\mathcal{F}$ in Figure \ref{fig6}b (see Table \ref{table1}). At times $t=40\,\mathrm{\omega_{ci}^{-1}}$ (black lines) and $t=87\,\mathrm{\omega_{ci}^{-1}}$ (red line) the spectra from the two runs are very similar, and correspond to a downstream thermal spread of $E_{th}\sim0.45\,\mathrm{E}_{up}$ ($\sim45\%$ of the incoming flow bulk energy). At this point in time, although there are non-thermal particles, the power law is very steep and the energy efficiency is very low. Measurements of the particle spectra at the same distance from the shock front at later times are shown in Figure \ref{fig6}a, and shows the power-law part of the distribution growing in energy as time elapses. This is a typical signature of particle acceleration due to a diffusive process. The complete fit to the particle spectrum at $t=400\,\mathrm{\omega_{ci}^{-1}}$ is shown in Figure \ref{fig6}c, and yields $E_{th}\sim0.43\,\mathrm{E_{up}}$, an injection energy of $E_{inj}\sim1.45\,\mathrm{E_{up}}$, and a power-law index of $\alpha\sim1.9$.
\begin{figure} 
\figurenum{15} 
\epsscale{0.6}
\plotone{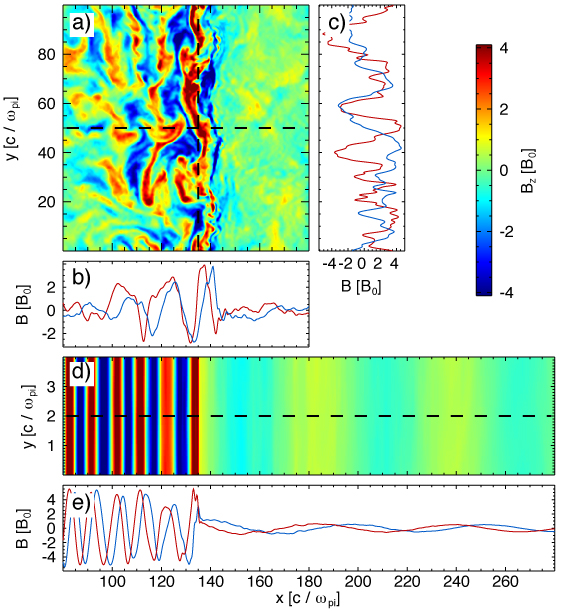}
\caption{Panel a) shows the $B_z$ magnetic field component for run $\mathcal{G}_3$.Panels b) and c) show lineouts of the $B_y$ (blue) and $B_z$ magnetic field components in the $x$ and $y$ directions along dashed lines. Frame d) shows the magnetic field for an identical run with a reduced transverse side (run $\mathcal{G}_4$ with $L_y=4\,\mathrm{c/\omega_{pi}}$), and frame e) shows the lineout for $B_y$ (blue) and $B_z$ for the same run.\label{fig14}} 
\end{figure}

Figure \ref{fig13} shows the global structure for the $M_A=4.7$ $\theta=0^\circ$ shock, simulated with the hybrid code, where a much larger $L_x=2000\,\mathrm{c/\omega_{pi}}$ $L_y=100\,\mathrm{c/\omega_{pi}}$ box was used. It is evident in the zoomed in box for the $B_z$ field component, Figure \ref{fig13}d, that there is a two-dimensional structure to the waves in the shock transition region. Also this wave structure is limited in size in the $x$ direction, and defines the region where the downstream plasma is not yet completely thermalized. If we now consider a scenario where the transverse simulation size is of the same order of magnitude of the upstream ion Larmor radius $r_{Li}=4.7\,\mathrm{c/\omega_{pi}}$, as it is implicitly done in 1D simulations, again a coherent and extended transition region is formed. This is shown in Figure \ref{fig14}, which compares the wave structures around the shock front for runs $\mathcal{G}_3$ ($L_y=100\,\mathrm{c/\omega_{pi}}$) and $\mathcal{G}_4$ ($L_y=4\,\mathrm{c/\omega_{pi}}$).

These results show that numerical simulation parameters, and in particular the simulation box sizes, must be chosen with care. The transverse box size must be at least comparable to the typical wavelength observed for waves propagating in the $x$ direction (parallel to the background magnetic field). To this requirement, we must add that the $x$ box size must be such as to include a significant fraction of the pre-shock, including the highest energy particles. This requirement translates into very long box sizes for quasi-parallel (or strictly parallel) shocks, which greatly increases the cost of running these simulations for long times $t>600\,\mathrm{\omega_{ci}^{-1}}$. Thus, for higher Mach numbers (increasing V, while keeping B constant) larger $x$ box sizes are needed, which justifies the $L_x=5000\,\mathrm{c/\omega_{pi}}$ and $L_x=8000\,\mathrm{c/\omega_{pi}}$ box sizes chosen for higher Mach numbers.

\bibliographystyle{apj}
\bibliography{bibliography}
\end{document}